# GRBs as cosmological probes—cosmic chemical evolution

**S Savaglio**
Max-Planck-Institut für Extraterrestrische Physik, Giessenbachstrasse, 85748 Garching, Germany
E-mail: savaglio@mpg.mpe.de

**Abstract.** Long-duration gamma-ray bursts (GRBs) are associated with the death of metal-poor massive stars. Even though they are highly transient events very hard to localize, they are so bright that they can be detected in the most difficult environments. GRB observations are unveiling a surprising view of the chemical state of the distant universe (redshifts $z > 2$). Contrary to what is expected for a high-$z$ metal-poor star, the neutral interstellar medium (ISM) around GRBs is not metal poor (metallicities vary from $\sim 1/10$ solar at $z = 6.3$ to about solar at $z = 2$) and is enriched with dust (90–99% of iron is in solid form). If these metallicities are combined with those measured in the warm ISM of GRB host galaxies at $z < 1$, a redshift evolution is observed. Such an evolution predicts that the stellar masses of the hosts are in the range $M_* = 10^{8.6-9.8} M_\odot$. This prediction makes use of the mass-metallicity relation (and its redshift evolution) observed in normal star-forming galaxies. Independent measurements coming from the optical–NIR (near-infrared) photometry of GRB hosts indicate the same range of stellar masses, with a typical value similar to that of the Large Magellanic Cloud (LMC). This newly detected population of intermediate-mass galaxies is very hard to find at high redshift using conventional astronomy. However, it offers a compelling and relatively inexpensive opportunity to explore galaxy formation and cosmic chemical evolution beyond known borders, from the primordial universe to the present.

**Contents**

## 1. Introduction

It is now widely accepted that long- or short-duration (longer or shorter than a few seconds) gamma-ray bursts (GRBs) are associated with the death of massive metal-poor stars and mergers of compact objects (neutron stars or black holes), respectively [1, 2]. GRBs are recognized as being potent cosmological tools [3], because they are detected at any distance (redshift), from $z = 0$ (the local universe) to the highest reached today at $z = 6.3$ [4], 860 million years after the Big Bang. They are so bright that they can 'illuminate' (for a short time) very 'dark' regions of distant galaxies, and probe the cosmic chemical evolution of the universe, from the most remote past, to today [5]–[8].

GRB spectra show features never observed before. The GRB light shines through a gas medium which is very different from the typical interstellar medium (ISM) observed with traditional tools in distant galaxies. In the past, observations of bright high-z QSOs (quasi-stellar objects) were used to explore the chemical enrichment in the universe [9, 10]. Strong absorption lines detected in QSO spectra, called damped Lyman-α (DLA) systems,[1] originate in galaxies crossing QSO sight lines (QSO-DLAs). Although QSOs are not transient sources, so that they can be observed at any time, they are not as bright as a typical GRB on its peak emission. The investigation of the very high redshift ($z > 6$) ISM with present-day technology using QSOs is rather complicated. In contrast, even though the 'weak' side of GRBs is their transitory nature, this is also their strength, because only transitory events can be so bright.

[1] The name originates from the strong Lyα absorption of the neutral hydrogen HI, which shows the natural damping profile for the high column densities ($> 10^{19}$ atoms $cm^{-2}$).

The fast fading of the source hampers observational programs, but the same limitation offers also the opportunity to identify and study in detail the host galaxy a few months later, when it is free of the bright GRB contamination. On the other hand, the detection of the emitting galaxy [11, 12] in QSO-DLAs is made virtually impossible by the overwhelming brightness of the background source, which is always present. Surprisingly, GRB hosts have often been proven to be intrinsically fainter than a typical galaxy at similar redshift, detected using traditional astronomy [13, 14]. The redshift of a faint and distant galaxy is hard to measure, whereas for GRB host galaxies the redshift comes as a bonus from the optical afterglow.

Recently, a cosmological scenario has become popular according to which massive galaxies formed abundantly in the remote past [15, 16] and turned off their star formation relatively quickly, whereas low-mass galaxies were active over a larger time interval [17, 18]. The chemical enrichment in the latter seems to have taken longer than in the former [19, 20]. GRB hosts, probing even lower stellar-mass galaxies [21] and spanning a large redshift interval, can be used very effectively to tune further theories of galaxy formation and evolution. In that regard, GRBs are still relatively unexplored and perhaps not sufficiently appreciated by the astronomical community.

In this contribution, we will describe the present knowledge of the chemical enrichment in GRB environments. Particular emphasis will be put on the metallicity measured through rest-frame UV absorption lines, and associated with the neutral and cold gas ($T \lesssim 10^3$ K) distributed along the GRB sight line. We call this the metallicity of the GRB-DLA. We also discuss the metallicity measured using emission lines, originating in the ionized warm gas ($T \sim 10^4$ K) of the starforming regions in the entire host galaxy. This is the host galaxy metallicity. The chemical enrichment measured from GRBs will be compared with the present knowledge of the cosmic chemical evolution, mainly derived using QSO-DLA studies, to establish what else GRBs can tell about this issue. As all GRBs discussed in the present study are classified as long-duration, we will not mention their duration for the remainder of the paper.

The paper is organized as follows: in section 2, we describe the technique used to measure heavy elements along a GRB sight line. In section 3, we will discuss the presence of dust, and how this affects our measurements. In section 4, we study the metallicity in GRB-DLAs as a function of redshift. Section 5 is on the metallicity measured in GRB hosts using emission lines. The detection of fine-structure features is described in section 6, while systematic effects are discussed in section 7. The final discussion is in section 8. Throughout the paper, we adopt a $h = H_o/100 = 0.7$, $\Omega_M = 0.3$, $\Omega_\Lambda = 0.7$ cosmology [22].

## 2. Chemical enrichment in GRB environments: the GRB-DLAs

Observations of UV absorption lines are further complicated by the steep GRB light curve. The advantage is that GRBs are (for some time) brighter than any other source in the universe and, if observed early on, numerous transitions associated with different elements can be detected relatively inexpensively. These are redshifted to the more accessible optical region for $z > 0.5$ (24 out of the 28 GRBs with measured redshift discovered by the GRB-dedicated mission Swift [23] are at $z > 0.5$). The relative abundances of different elements provide important clues on the physical state of the gas, its chemical enrichment and dust content, and the nucleosynthesis processes in the GRB starforming environment. As observed features resemble those of DLA systems in the high-$z$ ISM, detected in QSO spectra, these are called GRB-DLAs. To study them, we apply similar techniques [9, 10].

Column densities of neutral or singly ionized species (such as HI, NI, FeII, ZnII, SiII, i.e. with ionization potential above 13.6 eV) are often so high that the ionization correction is assumed to be negligible [24]. Hence, these column densities directly provide the column density of the corresponding element.[2] A fraction of some of these elements can be locked on to dust grains and would escape the UV detection. In these cases, to account for the total element column density, a dust depletion correction must be applied (see section 3). Alternatively, x-ray observations of the afterglow, though more difficult, provide the total column density [25, 26], because the absorption in x-ray energies is not sensitive to the element state (gas or solid).

Absorption lines provide the column density of elements, i.e. the number of atoms per unit surface responsible for the absorption. Because we lack the information on the geometry of the absorbing gas (i.e. the physical size parallel to the sight line over which the gas is distributed), the volume density of the gas cannot be directly measured. A large column density can equally indicate a high volume density of the gas, or a large physical size of the gas clouds. However, relative abundances (including metallicity) do not suffer from this limitation.

The Doppler parameter (line width) is in principle related to the temperature or the level of turbulence in the gas. In practice, the GRB sight line is crossing a complex physical structure with multiple absorptions, not resolved by the poor spectral resolution often used. In these cases, a single absorption line is observed whose width does not carry information on the detailed physical distribution of the gas, therefore it is called 'effective' Doppler width [27, 28].

The best way to derive column densities is by fitting the absorption lines in high resolution/signal-to-noise (S/N) spectra with a Voigt profile. When the GRB is too faint and only low-resolution spectroscopy is possible, other techniques must be used. In the case of high S/N, the curve of growth (COG) analysis [29] is robust for weak lines. For instance, the linear approximation can be applied for equivalent widths $\mathrm{EW} < 0.2$ Å and effective Doppler parameter $b > 30\ \mathrm{km\ s^{-1}}$. For stronger lines, the COG gives reliable results if several lines with different oscillator strengths and EWs of the same ion are detected. As an alternative approach, the apparent optical depth (AOD) method [30] can be applied to medium-resolution and reasonable S/N spectra, for weak to moderately strong lines. The most difficult scenario is when the S/N is poor ($< 10$). Regardless of resolution, any technique can lead to misleading results, because only heavily saturated lines, which fall in the flat part of the COG, can be detected.

In table 1, we report HI and heavy element column densities, plus errors when available, measured in 22 GRB afterglows, i.e. more than 1/4 of all GRBs with measured redshift (figure 1). The column densities are derived using all methods described above (except the AOD) and results are published in 26 different papers (including the present one). As the sample is still very limited, and errors are often large, any statistical analysis of physical parameters and their evolution with redshift is affected by large uncertainties. GRB-DLA science is still at the dawn with respect to QSO-DLA studies, for which high resolution spectroscopy was acquired for $\sim$200 targets. On the other hand, the advent of the GRB mission Swift [23], together with the fast development of dedicated observational programs, will soon provide us with a competitive sample.

The HI, FeII and SiII column densities are measured in 15, 14 and 9 GRB-DLAs, respectively. From a quick comparison with the hydrogen and heavy elements seen in QSO-DLAs (figure 2), one immediately recognizes that GRB-DLAs are generally characterized by larger column densities.

[2] Higher ionized species (e.g. CIV, SiIV) often detected at the same redshift of the low-ionization species, are very likely associated with different regions along the GRB sight line, such as the hot-ionized gas in the halo of the host, and has little to do with the GRB-DLA.

**Table 1.** GRB-DLA sample.

| GRB | z | log $N_{HI}$ | log $N_{FeII}$ | log $N_{ZnII}$ | log $N_{CrII}$ | log $N_{MnII}$ |
|---|---|---|---|---|---|---|
| 990123 | 1.6004 | – | $14.78^{+0.17}_{-0.10}$ | 13.95 ± 0.05 | – | – |
| 000301C | 2.040 | 21.2 ± 0.5 | – | – | – | |
| 000926 | 2.038 | 21.3 ± 0.2 | $15.6^{+0.20}_{-0.15}$ | 13.82 ± 0.05 | 14.34 ± 0.05 | – |
| 010222 | 1.475 | – | $15.32^{+0.15}_{-0.10}$ | 13.78 ± 0.07 | $14.04^{+0.04}_{-0.06}$ | $13.614^{+0.08}_{-0.06}$ |
| 011211 | 2.142 | 20.4 ± 0.2 | 14.7 ± 0.2 | – | – | – |
| 020124 | 3.198 | 21.7 ± 0.2 | – | – | – | – |
| 020405 | 0.691 | – | 15.05 ± 0.75 | – | – | – |
| 020813 | 1.255 | – | 15.48 ± 0.04 | 13.54 ± 0.06 | 13.95 ± 0.03 | 13.62 ± 0.03 |
| 021004 | 2.33 | 19.5 ± 0.5 | $13.40^{+0.14}_{-0.13}$ | – | – | – |
| 030226 | 1.986 | – | > 14.8 | – | – | – |
| 030323 | 3.371 | 21.90 ± 0.07 | 15.93 ± 0.08 | < 14.5 | – | – |
| 030328 | 1.522 | – | $14.3^{+0.6}_{-0.2}$ | – | – | – |
| 030429 | 2.658 | 21.6 ± 0.2 | – | – | – | – |
| 050401 | 2.8992 | 22.6 ± 0.3 | 16.0 ± 0.2 | 14.4 ± 0.2 | 14.6 ± 0.2 | – |
| 050505 | 4.275 | 22.05 ± 0.10 | 15.5 | – | – | – |
| 050730 | 3.9685 | 22.15 ± 0.05 | 15.35 | – | – | – |
| 050820 | 2.6147 | 21.0 | 15.4 | – | – | – |
| 050904 | 6.295 | 21.62 ± 0.20 | – | – | – | – |
| 050908 | 3.3437 | 19.2 | – | – | – | – |
| 051111 | 1.5495 | – | $15.22^{+0.09}_{-0.08}$ | 13.58 ± 0.15 | 13.85 ± 0.1 | 13.52 ± 0.15 |
| 060206 | 4.048 | 20.85 ± 0.10 | – | – | – | – |
| 060522 | 5.11 | 20.5 ± 0.5 | – | – | – | – |

Results derived using the COG have been recently questioned [50], based on the analysis of the spectrum of GRB 051111, where the COG results are compared with the Voigt profile fitting. In subsection 7.1, we present a similar test using the good-quality Keck spectrum of the high redshift starforming galaxy cB58, at $z = 2.72$ [51]. The AOD and the COG give consistent results.

### 3. Dust properties

Some heavy elements in the cold ISM are partly locked on to dust grains. Column densities measured using UV absorption lines account for the gas phase only, and the fraction in dust escape the detection. In the ISM of the Milky Way, the fraction of iron in gas phase ranges from about 20% of the total in the warm and diffuse gas of the halo, to less than 1% in the cool disc ISM [52]. On the other hand, the fraction of zinc in grains is generally negligible, except in the cool disc ISM, where it can be as high as 20% [52]. Therefore, when measuring heavy element abundances, one has to consider dust correction for some of them.

The dust content in a GRB-DLA can be characterized by comparing abundances of elements with different depletion characteristics. The iron-to-zinc relative abundance is the best diagnostic parameter, because these two elements are easier to detect than others, and they are also quite extreme in their refractory properties. Moreover, because they are both iron-peak elements with

**Table 1.** (continued)

| GRB | log $N_{SiII}$ | log $N_{SiII*}$ | log $N_{SII}$ | log $N_{NiII}$ | Reference |
|---|---|---|---|---|---|
| 990123 | – | – | – | – | [6] |
| 000301C | – | – | – | – | [31] |
| 000926 | $16.47^{+0.10}_{-0.15}$ | – | – | – | [6, 32] |
| 010222 | 16.09 ± 0.05 | – | – | – | [6] |
| 011211 | $15.1^{+0.6}_{-0.3}$ | – | – | – | [33] |
| 020124 | – | – | – | – | [34] |
| 020405 | – | – | – | – | [35], this study |
| 020813 | 16.29 ± 0.04 | 14.32 ± 0.11 | – | 14.20 ± 0.03 | [36] |
| 021004 | – | – | – | – | [37, 38] |
| 030226 | > 14.7 | – | – | – | [39] |
| 030323 | $16.15^{+0.26}_{-0.19}$ | 14.182 ± 0.036 | 15.84 ± 0.19 | – | [7], this study |
| 030328 | < 16.0 | – | – | – | [40] |
| 030429 | – | – | – | – | [41] |
| 050401 | 16.5 ± 0.4 | – | – | – | [25] |
| 050505 | 15.7 | 15.1 | 16.1 | 14.6 | [42] |
| 050730 | – | – | 15.42 ± 0.09 | – | [28, 43, 44] |
| 050820 | 15.95 | – | – | – | [45] |
| 050904 | < 16.6 | 14.0 ± 0.4 | 15.95 ± 0.55 | – | [4, 46], this study |
| 050908 | – | – | – | – | [47] |
| 051111 | 16.18 ± 0.2 | 14.96 ± 0.10 | – | 13.99 ± 0.35 | [48, 49] |
| 060206 | > 14.56 | 14.42 ± 0.02 | 15.21 ± 0.03 | – | [8] |
| 060522 | – | – | – | – | [50] |

Notes: High spectral resolution (FWHM $\lesssim$ 10 kms) and reasonable signal-to-noise ratio (S/N > 10) were obtained for GRB 050820, GRB 050730, GRB 051111, and GRB 060522.

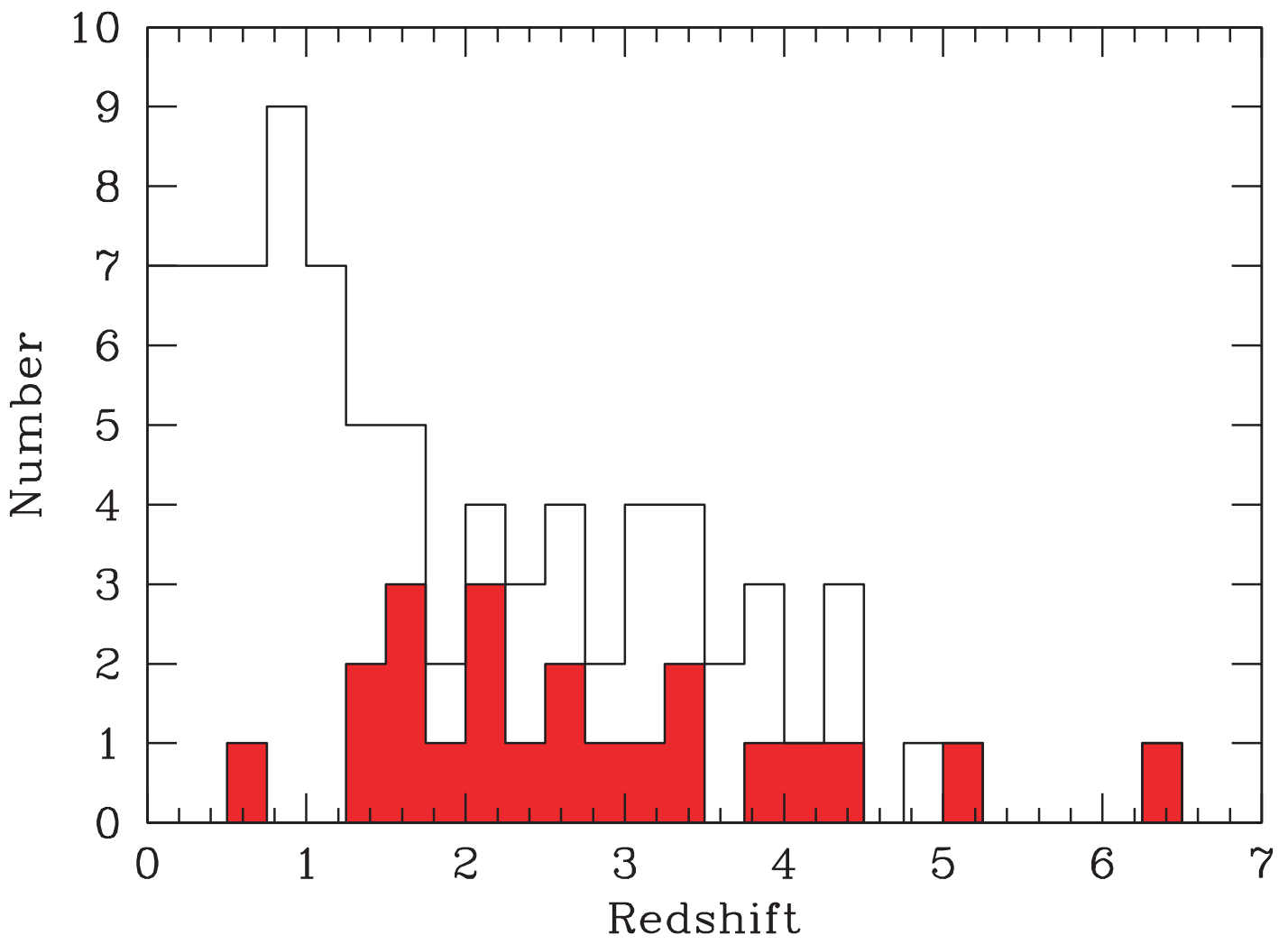

**Figure 1.** Number of GRBs per unit redshift interval for the total sample with known redshift (82 objects, *empty histogram*) and for those studied here and listed in table 1 (22 objects, *filled histogram*).

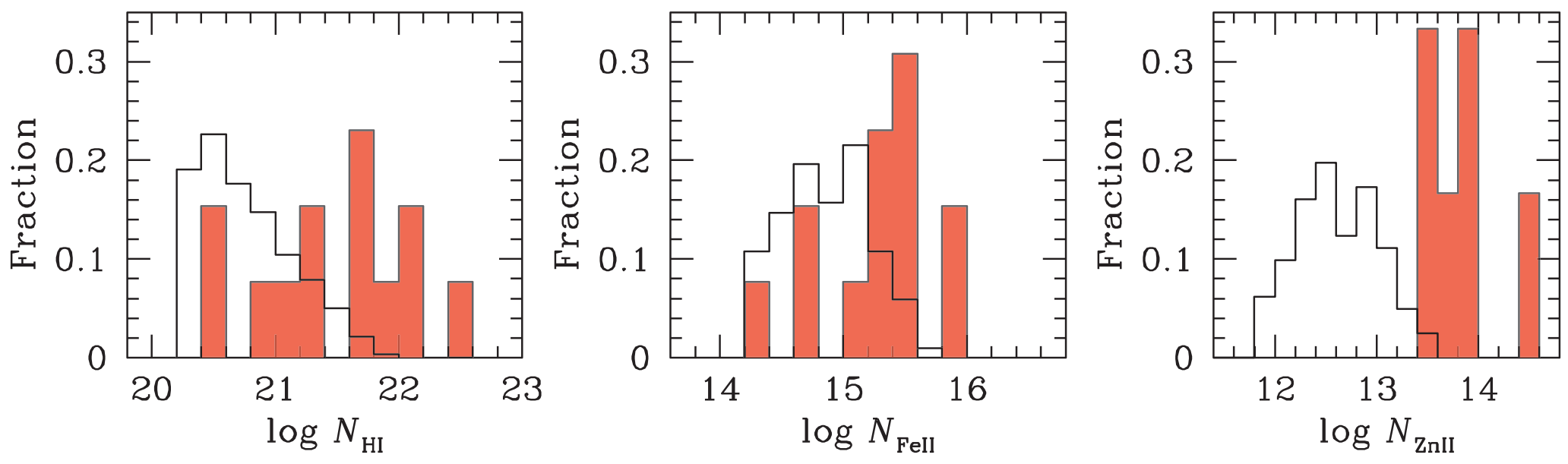

**Figure 2.** Fraction of GRB-DLAs (*filled histograms*) and QSO-DLAs (*empty histograms*) per HI, FeII and ZnII column-density interval (left-, middle- and right-hand side panels, respectively). The QSO-DLA histograms are complete for $\log N_{\rm HI} \geqslant 20.2$, $\log N_{\rm FeII} \geqslant 14.2$ and $\log N_{\rm ZnII} \geqslant 11.8$. It is apparent that column densities in GRB-DLAs are generally higher than in QSO-DLAs. This can either indicate that GRB-DLAs originate in bigger galaxies, or that the volume density of the gas in GRB-DLAs is higher, or both.

similar formation timescales ([53] and references therein), it is reasonable to assume that if their relative abundance in the ISM is very different than in stars, the missing fraction is locked in dust.

From table 1, we derive the iron-to-zinc relative abundance [Fe/Zn] in six GRB-DLAs[3] and plot the results in figure 3, as a function of the ZnII column density. In figure 3, we also show the same quantities for 59 QSO-DLAs (redshift interval $z = 0.6$–$3.4$), and detections in the local ISM. GRB-DLAs and QSO-DLAs occupy two distinct regions in the plot, suggesting that they belong to two distinct populations. The mean [Zn/Fe] value in the QSO-DLA sample is $-0.46$, and the two parameters are correlated at $5\sigma$ level (more than 99.998% significant). In the six GRB-DLAs, [Fe/Zn] ranges from $-1$ to $-2$. If [Fe/Zn] measures the dust depletion, figure 3 indicates that iron in GRB-DLAs is 90 to 99% locked into dust grains. If we consider the two samples together (QSO-DLAs and GRB-DLAs, ignoring error bars), the two parameters are correlated at $7.3\sigma$ level, and the linear fit is:

$$[{\rm Fe/Zn}] = (-0.494 \pm 0.068) \log N_{\rm ZnII} + (5.73 \pm 0.85). \qquad (1)$$

This correlation is confirmed by measurements in the ISM of the Milky Way and Magellanic Clouds.

The simplest interpretation of this correlation is that the dust depletion (or equivalently the dust-to-metals ratio) tends to be higher for higher ZnII column densities. It is not clear why this should be the case. The ZnII column density is on average higher for higher volume density of zinc, and that would give a more favourable ambient for the creation, or the non-destruction, of dust grains in the medium. On the other hand, $N_{\rm ZnII}$ can also be higher when the cross-section of the gas responsible for the absorption is higher (e.g. for larger gas clouds and/or a high number of clouds along the sight lines, which is possible for bigger galaxies) regardless of the volume density. Better statistics (more GRB-DLA observations) will certainly help in clarifying this issue. For a detailed discussion of the dust grain properties (creation, destruction, chemical composition) in the ISM of the Milky Way, see [54].

[3] By definition, $[{\rm Fe/Zn}] = \log(N_{\rm FeII}/N_{\rm ZnII}) - \log({\rm Fe/Zn})_\odot$, valid for negligible ionization correction. Here, [Fe/Zn] is measured for the gas phase only. If $[{\rm Fe/Zn}] = -1, -2$, this suggests that approximately 10% or 1% of iron is in gas form, respectively, and the rest is in solid form.

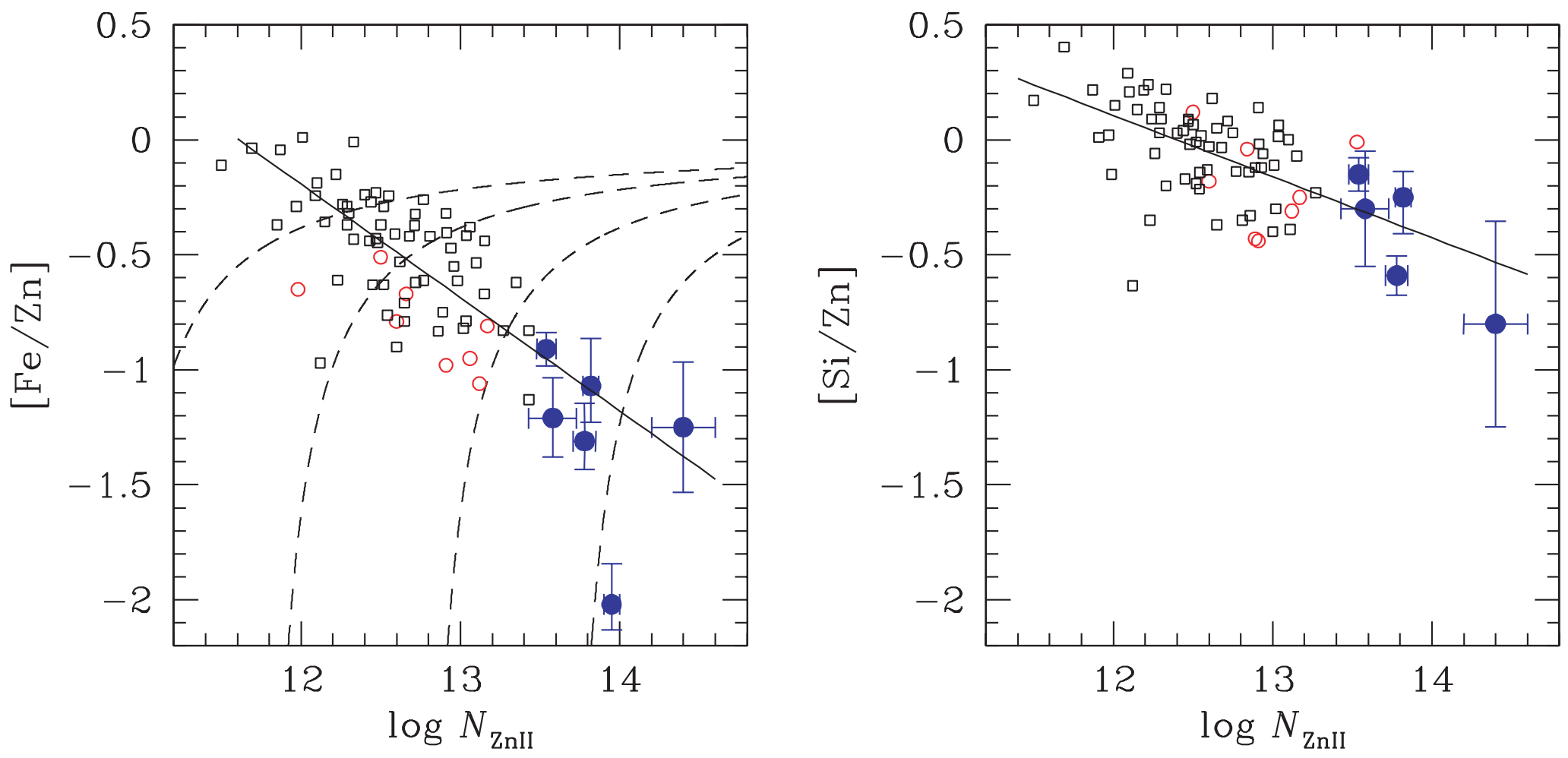

**Figure 3.** Iron-to-zinc (left-hand side panel) and silicon-to-zinc (right-hand side panel) relative abundances as a function of ZnII column density. GRB-DLAs (●) and QSO-DLAs (□) occupy two distinct regions in the plots, suggesting a different physical nature for these two populations. The [Fe/Zn] and [Si/Zn] versus $N_{\mathrm{ZnII}}$ correlations (solid lines) for the GRB-DLAs and QSO-DLAs samples together are 7σ and 5.7σ significant, respectively. (○) are measurements from the ISM of the Milky Way and Magellanic Clouds (see [6] and references therein). The dashed curves in the left-hand side panel are model predictions expressed by equations (5) and (6) for a constant visual extinction, from left to right $A_V = 0.005, 0.05, 0.5, 5$, and for $a = 0.5$ (see text).

When only ZnII is measured, the expected [Fe/Zn] value can be estimated using equation (3), with an uncertainty of typically a factor of 2.5 (the [Fe/Zn] versus log $N_{\mathrm{ZnII}}$ distribution in figure 3 has a 1σ dispersion of 0.4 dex).

The α same behaviour is observed when silicon and zinc are considere (figure 3) but the slope is not as steep as for iron. The [Si/Zn] versus log $N_{\mathrm{ZnII}}$ linear fit is:

$$[\mathrm{Si/Zn}] = (-0.266 \pm 0.047) \log N_{\mathrm{ZnII}} + (3.30 \pm 0.60). \quad (2)$$

The correlation between $N_{\mathrm{FeII}}$ and $N_{\mathrm{ZnII}}$ can be described by the following bisector fit:

$$\log N_{\mathrm{FeII}} = (0.679 \pm 0.053) \log N_{\mathrm{ZnII}} + (6.37 \pm 0.67), \quad (3)$$

or its inverse:

$$\log N_{\mathrm{ZnII}} = (1.47 \pm 0.11) \log N_{\mathrm{FeII}} + (-9.4 \pm 1.7), \quad (4)$$

and are only valid for $\log N_{\mathrm{FeII}} > 14$ or $\log N_{\mathrm{ZnII}} > 12$.

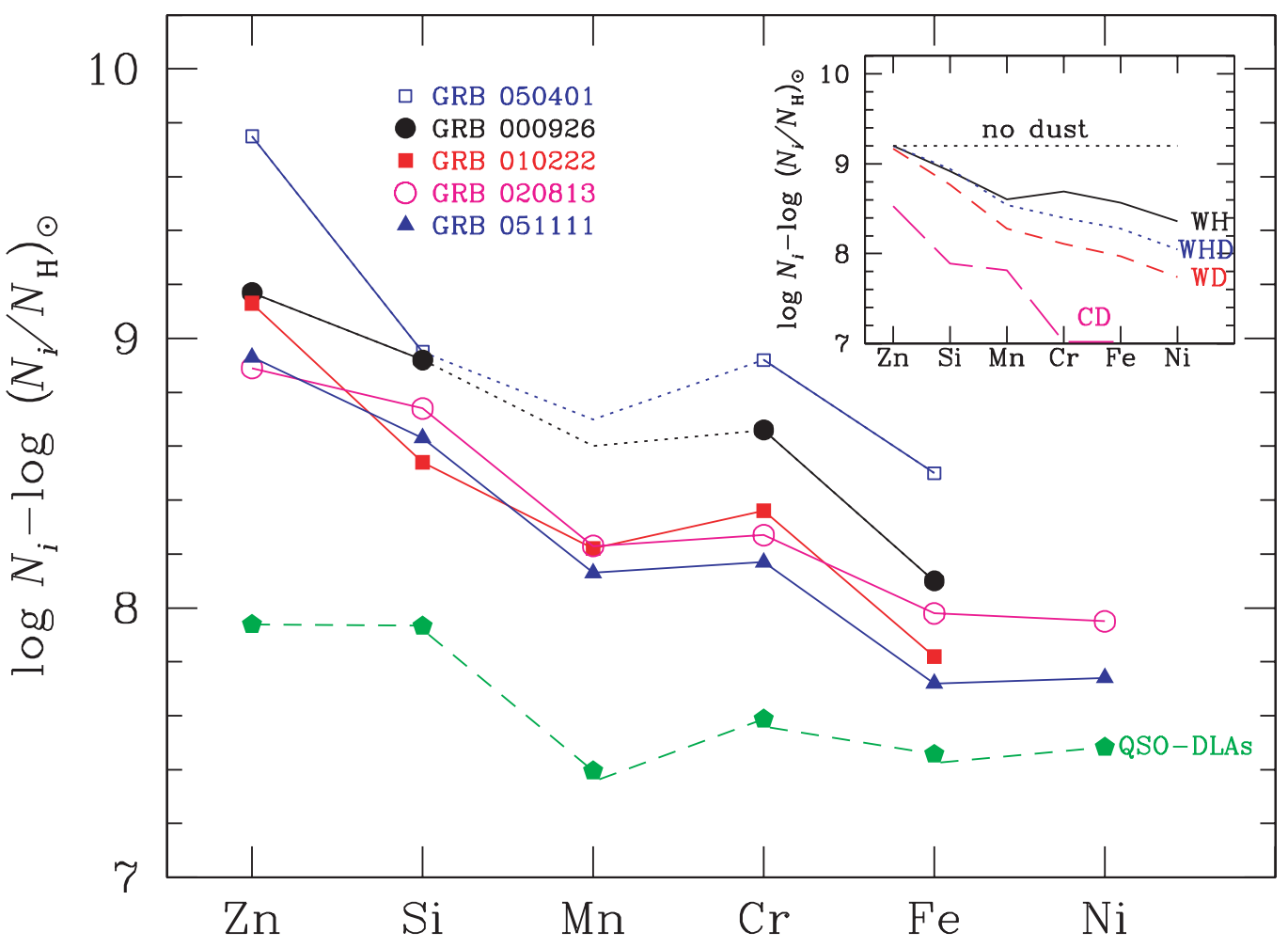

**Figure 4.** Depletion pattern in five GRB-DLAs with four or more heavy elements measured (errors are $\leqslant$0.2 dex). The Mn column density is not measured in GRB 000926 and GRB 050401; expected values are suggested by the dotted lines for display purposes only. The dashed line at the bottom is the mean depletion pattern measured in 20 QSO-DLAs for which all six elements are detected. Metal column densities and dust depletion in GRB-DLAs are much higher than in QSO-DLAs. For instance the zinc-to-iron relative abundance (representing dust depletion) is on average four times larger in GRB-DLAs ([Zn/Fe] $\simeq$ 1.1) than in QSO-DLAs ([Zn/Fe] $\simeq$ 0.5). In the inset, the dust depletion patterns (rescaled to an arbitrary value) of the Galactic ISM are plotted for comparison [52]; from top to bottom: warm halo, warm halo + disc, warm disc, cool disc clouds.

### *3.1. Dust-depletion patterns*

The dust depletion in the ISM of the Milky Way for different heavy elements shows patterns that are a function of the gas physical environment, such as density and temperature. The dust depletion is larger for denser and cooler clouds [52], as shown in the inset of figure 4. Dust depletion patterns derived for QSO-DLAs [55] (dashed line in figure 4) are more similar to the depletion pattern of the Galactic warm halo.

We investigate this issue for the five GRB-DLAs for which the column density of at least four elements is estimated. Results in figure 4 resemble the depletion patterns observed in the Milky Way, in particular those of the warm disc + halo and warm disc gas (inset in figure 4). This similarity strongly supports the existence of dust in GRB-DLAs.

We note that there is no indication of zinc depletion. If zinc were depleted, like in the cool gas of the Milky Way, we would expect [Mn/Cr] $\sim$ 0.9, whereas [Mn/Cr] $\sim$ 0 is observed. However, such a possibility cannot be excluded. A large Zn dust depletion would solve the large oxygen-to-zinc relative abundances found in the x-ray spectra of two GRB afterglows [25, 26].

We assume for the remainder of the paper that $N_{\mathrm{ZnII}}$ measured in UV spectra gives a good indication of the total zinc column density (the zinc in solid phase is neglected). More observations are necessary to confirm the validity of this assumption.

### 3.2. Dust extinction

The GRB emission is extincted by the dust distributed along its sight line. Partly, the dust is in the Galaxy (this can be easily taken into account), partly in the intervening material between the GRB and the Galaxy (typically negligible, but see [56, 57]), and partly in the vicinity of the GRB or in the host galaxy, across several kpc.

The dust extinction in the GRB environment is directly proportional to the column of dust. It is reasonable to assume that this is proportional to the total column of metals (for instance represented by the ZnII column density) and to the dust depletion (for instance represented by [Fe/Zn]). It is also reasonable, to first order, to assume that the dust extinction is not directly proportional to the quantity of HI distributed along the sight line, so it can be estimated regardless of whether the HI column density is known or not.

The dust extinction law $A_V$ is expressed as $f_{int} = f_{obs} 10^{-0.4 A_V}$, where $f_{int}$ and $f_{obs}$ are the intrinsic and dust-extincted GRB-afterglow fluxes, respectively. The dust extinction is commonly called reddening (because the UV is more absorbed than the optical) and is often expressed in terms of extinction in the visual band $A_V$. At high redshift, the overall extinction law is assumed to be as in the Milky Way or in the Large and Small Magellanic Clouds (LMC and SMC), the only well-known extinction laws. It is worth recalling that not only the extinctions are very different in these three galaxies, but also that large deviations from the 'standard' laws are observed within the same galaxy [58, 59]. This means that correcting the GRB observed flux using the extinction laws known for the local universe, though it is the best that can be done at this time, is only a rough approximation that needs to be tested through different means [60].

We use a simpler approach and estimate the visual extinction $A_V$ by comparing the amount of metals distributed along the GRB sight line with the same quantities observed in the local universe, for which $A_V$ is measured. We consider, as a reference point, the $A_V$ measured in the ISM of the LMC, but the difference with the Milky Way or SMC is not very large. Such a normalization is more appropriate than others, because the LMC is a small starforming galaxy with subsolar metallicity, probably close to a typical GRB host galaxy. The rate of visual extinction in the LMC is usually indicated as $A_V^{LMC} = 0.4$ for a column of HI of $10^{21}$ cm$^{-2}$ [61]. As the LMC metallicity is about 1/4 solar, this can be translated into $A_V = 0.4$ for a ZnII column of $\log N_{ZnII} = 13.1$. We also consider that in the LMC the dust depletion is not negligible, and the mean iron-to-zinc abundance is $[Fe/Z] \simeq -0.9$ [62].

If we assume that in GRB-DLAs, zinc is not depleted on to dust grains, then $[Fe/Zn] \sim 0$ means $A_V \sim 0$ (no dust). The visual extinction can also be considered negligible when the column density of ZnII is relatively small, regardless of the dust-to-metals ratio. For instance, for an LMC dust-to-metals ratio, $A_V < 0.01$ when $\log N_{ZnII} < 12$. It is also reasonable to assume that the larger the column of metals (represented by $N_{ZnII}$), the higher the extinction. However, these two quantities are not just directly proportional, but extinction saturates for metal column densities above some threshold. Clumpy dust (in contrast to diffuse dust), important for high gas density, could give saturation [63].

We express $A_V$ in terms of the column of dust represented by [Fe/Zn], as:

$$A_V = A_V^{LMC} N_{ZnII} 10^{\{a/[Fe/Zn]+b\}} \tag{5}$$

which is valid only for $[Fe/Zn] < 0$, and where $a$ and $b$ are constant. If we normalize to the LMC $A_V$, $a$ and $b$ are related by the following expression:

$$b = -13.1 + \frac{a}{0.9} \tag{6}$$

As an example, in figure 3, we show the case for $a = 0.5$. The value of $a$ is higher for a more important clumpiness of the dust in the medium.

This approach gives an estimate of $A_V$ when FeII and ZnII column densities are measured and is a simplification of the method used for GRB 020813 [27]. When only ZnII is available, then [Fe/Zn] can be assumed using equation (1). In order to estimate the dust extinction for different wavelengths, an additional assumption on the extinction law has to be made.

Given the large metal column densities generally measured in GRB-DLAs, one expects a high dust extinction. In contrast, GRB afterglows often show a featureless power-law behaviour in the rest-frame UV [7, 43, 64]. This lack of reddening is at odds with the presence of a possible high dust depletion. One way out is the already mentioned high clumpiness of the dust, i.e. an ISM polluted by pockets of dust by which the GRB emission is hardly affected [63].

Alternatively, the extinction law may be a weak function of the wavelength, which is possible if the GRB emission is preferentially destroying small size grains [65]–[68]. The GRB afterglow radiation can certainly readjust the grain-size distribution. On the other hand, as small grains evaporate during the process, bigger grains could be broken in small grains, therefore the total effect might not be as efficient as foreseen.

A flatter extinction law in the rest-frame 2800–1800 Å range is predicted for QSOs or SNII environments [69]. If a similar extinction is affecting GRBs, the reddening would be small for observed optical spectra in the redshift range $z = 0.7$–$2.5$.

## 4. GRB-DLA metallicities and redshift evolution

Measuring the metal content in the cold gas distributed along the GRB sight line (i.e. the GRB-DLA), is further complicated by the limited access to the hydrogen column density. Indeed, no instruments are available at this time to measure the metallicity in GRB-DLAs at $z < 1.6$, because the Ly$\alpha$ absorption line is still in the UV.[4] More than half of all GRBs with measured redshift are at $z < 1.6$.

Nonetheless, table 1 shows that HI and one or more heavy elements are detected in several GRB-DLAs, for which in principle the metallicity (the total metal content relative to the total hydrogen) can be estimated. As described in section 3, some heavy elements are partly locked on to dust grains, and for those a dust depletion correction is necessary (see figure 4). Our metallicity estimate in these GRBs is done in the following way:

1. The column of hydrogen can be approximated by the HI column density. This is directly estimated from the Ly$\alpha$ absorption line.
2. When ZnII is measured (GRB 000926 and GRB 050401), then the metallicity is simply given by [Zn/H], with no dust correction.

[4] STIS on board of the Hubble Space Telescope, the most advanced UV spectrograph, stopped working in August 2004. The next suitable instrument, COS, will be installed on HST during the next servicing mission, not before the end of 2007.

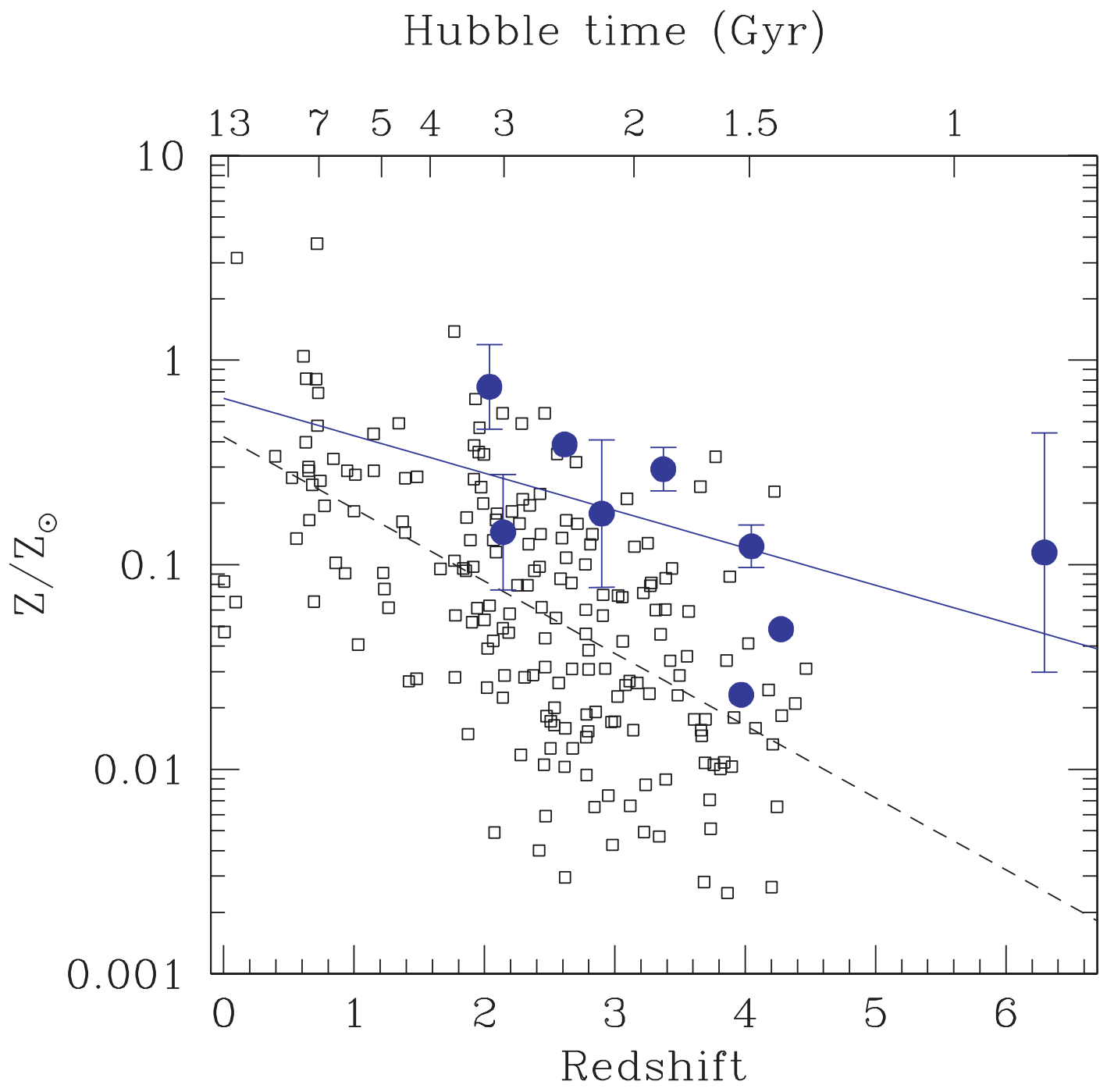

**Figure 5.** Redshift evolution of the metallicity relative to solar values, for nine GRB-DLAs (●) and 197 QSO-DLAs (□). Error bars are not available for three GRB-DLAs. Errors for QSO-DLAs (measured from the element column density uncertainties) are generally smaller than 0.2 dex. The solid and dashed lines indicate the best-fit linear correlation for GRB-DLAs and QSO-DLAs, respectively. The GRB-DLAs metallicity is on average ∼5 times larger than in QSO-DLAs. The upper horizontal *x*-axis indicates the age of the Universe (Hubble time).

3. For those GRBs with no ZnII detection, FeII is used (errors are generally smaller than for SiII), and dust-depletion correction is estimated using the $N_{ZnII}$ versus $N_{FeII}$ correlation expressed by equation (4), valid for $\log N_{FeII} > 14$.[5]
4. If neither SiII nor FeII are measured, then SII is used, with no dust depletion correction.

Results for nine GRB-DLAs are in figure 5 and table 2. Despite the large uncertainties which, by the way, take into account only the error in the column density measurements, the sample shows a weak redshift evolution. The linear correlation is described by (errors are not considered):

$$\log Z/Z_\odot = (-0.18 \pm 0.10)z + (-0.19 \pm 0.34). \qquad (7)$$

A similar study was conducted by [41] and [8] (six and nine GRBs were used, respectively) and in both papers a redshift evolution of GRB-DLA metallicities is suggested. Differences with equation (7) are mainly due to the dust depletion correction applied by us.

[5] GRB 021004 is excluded for its complex absorption structure [36, 70] and for the small column density of HI and FeII, which might indicate a non-negligible ionization correction.

**Table 2.** GRB-DLA metallicities.

| GRB | z | [S/H] | [Zn/H] | [Si/H] | [Fe/H] | $\log Z/Z_{\odot}$ [a] (adopted) | Ion[b] |
|---|---|---|---|---|---|---|---|
| 000926 | 2.038 | – | $-0.13 \pm 0.21$ | $-0.38^{+0.22}_{-0.25}$ | $-1.2^{+0.28}_{-0.25}$ | $-0.13 \pm 0.21$ | ZnII |
| 011211 | 2.142 | – | – | $-0.85^{+0.63}_{-0.36}$ | $-1.20 \pm 0.28$ | $-0.84 \pm 0.28$ | FeII |
| 030323 | 3.371 | $-1.33 \pm 0.20$ | $< -0.05$ | $-1.30^{+0.27}_{-0.20}$ | $-1.47 \pm 0.11$ | $-0.53 \pm 0.11$ | FeII |
| 050401 | 2.899 | – | $-0.75 \pm 0.36$ | $-1.55 \pm 0.50$ | $-2.00 \pm 0.36$ | $-0.75 \pm 0.36$ | ZnII |
| 050505 | 4.275 | $-1.22$ | – | $-1.9$ | $-2.0$ | $-1.3$ | FeII |
| 050730 | 3.969 | $-2.00 \pm 0.10$ | – | – | $-2.3$ | $-1.6$ | FeII |
| 050820 | 2.615 | – | – | $-0.60$ | $-1.1$ | $-0.4$ | FeII |
| 050904 | 6.295 | $-0.94 \pm 0.59$ | – | $< -0.6$ | – | $-0.94 \pm 0.59$ | SII |
| 060206 | 4.048 | $-0.91 \pm 0.10$ | – | $> -1.8$ | – | $-0.91 \pm 0.10$ | SII |

[a]Only errors on column densities are considered.
[b]Ion used to estimate metallicity.

In figure 5 we include the metallicities estimated for QSO-DLAs. These are derived in the following way: if ZnII is measured (82 objects), the metallicity is given by the zinc-to-hydrogen relative abundance (assuming that zinc is not depleted, as indicated by the observed dust depletion patterns). When Zn is not measured, SiII is used (88 objects), and the correction for dust depletion is done using the mean silicon-to-zinc relative abundances measured in 56 QSO-DLAs, [Si/Zn] = − 0.03 with a dispersion of 0.19. When both Zn and Si are not measured, FeII is used (27 objects), and the correction for dust depletion is done using the mean iron-to-zinc relative abundances measured in 66 QSO-DLAs, [Fe/Zn] = − 0.48 with a dispersion of 0.27. The slope of the redshift evolution for QSO-DLAs is $\Delta \log Z/\Delta z = -0.352 \pm 0.036$ (dashed line in figure 5).

Despite the poor number of statistics, GRB-DLAs tend to be more metal-rich than QSO-DLAs (on average by a factor of 5) and to have a flatter redshift evolution. A substantial difference between QSO and GRB absorption lines in general was recently detected [56, 57]. In particular, MgII absorption systems intersecting GRB sight lines (intervening MgII systems) are stronger than in QSO MgII systems. Given the small sample, any conjectures need the support of more data. However, GRB studies revealed a new picture of the gas properties in the high redshift universe, with large metal column densities never observed before. GRB-DLAs and QSO-DLAs belong to two different classes of absorbers, although they both trace the ISM in high-z galaxies.

One has to consider that the GRB environment can be investigated thanks to the fact that a bright $\gamma$–ray emission is detected, and this is not preselected according to luminosity or colour of the host galaxy (like most galaxy surveys), or of the brightness of the background source (as for QSO-DLAs). So, to first order, GRB-DLAs are less affected by observational limitations. On the other hand, GRBs are occurring in regions of star formation [71] for which metal enrichment can be faster than in the average galaxy ambient. It was recently suggested that, because of their low metallicity, QSO-DLAs probe preferably low-mass galaxies [19], because less massive galaxies tend to be more metal-poor [72]. GRB-DLAs, in contrast, might be associated with more massive galaxies, on average as massive as the LMC [21]. Therefore, perhaps it is not surprising that the metallicity found through GRB studies is higher than in QSO-DLAs.

The cosmic metal enrichment probed by GRB-DLAs needs a larger sample. This field will expand very quickly, because Swift is providing a large number of redshifts, especially in the $z > 1.5$ regime, where optical spectroscopy is possible. The main limitation is now the low redshift regime (below $z = 1.5$ when the universe was only 4 Gyr old) because UV spectroscopy is necessary, and at this time no instruments are available for such studies.

The metallicity in GRB-DLAs can be complemented with metallicities estimated from the emission lines in the host galaxies (the warm ionized gas). These are particularly important because they can be derived in the optical for $z < 1$ targets. At larger redshift galaxies are fainter, plus suitable emission lines are redshifted to the NIR, where spectroscopy is harder to acquire. This will be discussed in section 5.

### *4.1. The $\alpha$ elements*

The element formation in an environment where a GRB takes place is expected to carry the signature of nucleosynthesis processes typical for massive metal-poor stars [53, 75, 76]. Relative abundances of massive metal-poor stars are, for instance, characterized by an overabundance of $\alpha$ elements[6] such as O, Si, S, Ca and Mg.

In GRB-DLAs, the $\alpha$ elements observed are S and Si. Sulfur is often referred to as a good element to estimate metallicity in general, because in the Galactic ISM it is not affected by dust depletion. Here, the silicon-to-sulfur relative abundance ranges from $[S/Si] = 0.3$ to 1.3 from the warm halo to the cool dense clouds [52].

In GRB-DLAs, sulfur is measured in a few objects (figure 6 and table 1), and it can be used to derive metallicity. In QSO-DLAs, surprisingly, the sulfur relative abundances are lower than expected, even considering the generally low dust depletion. For instance, the mean sulfur-to-silicon abundance is $[S/Si] = -0.05$, whereas an average of $\sim$0.2–0.3 is expected.

The same effect seems to be present in GRB-DLAs (figure 6), though the statistics are still too poor to draw any conclusion. The SII column density can be underestimated if the effective Doppler parameter is overestimated [50]. Also, a poor deblending of the absorption lines, often contaminated by the wing of the Ly$\alpha$ profile or by the strong SiII$\lambda$1260 absorption, can lead to underestimation of the SII column density. Another danger is a wrong assumption on the oscillator strengths of the lines.

The situation seems different for [Fe/Si], where a low negative value can indicate both an $\alpha$ element overabundance and/or a large Fe dust depletion with respect to Si. At this stage, nothing definitive can be concluded and it is necessary to have more data at high S/N and resolution.

## 5. GRB host metallicities

In contrast to what is generally possible with QSO-DLAs [42], the GRB host galaxy can be observed when the bright emission of the GRB is gone, a few days after the burst. Often the ionized gas (the HII regions), pumped by the UV photons coming from the intense star formation, is emitting with sufficient intensity so emission lines can be detected, even if the galaxy is faint and small. Therefore a study of the GRB host metallicity is efficiently performed.

[6] The $\alpha$ elements have atomic numbers multiple of the atomic number of the $\alpha$ particle He.

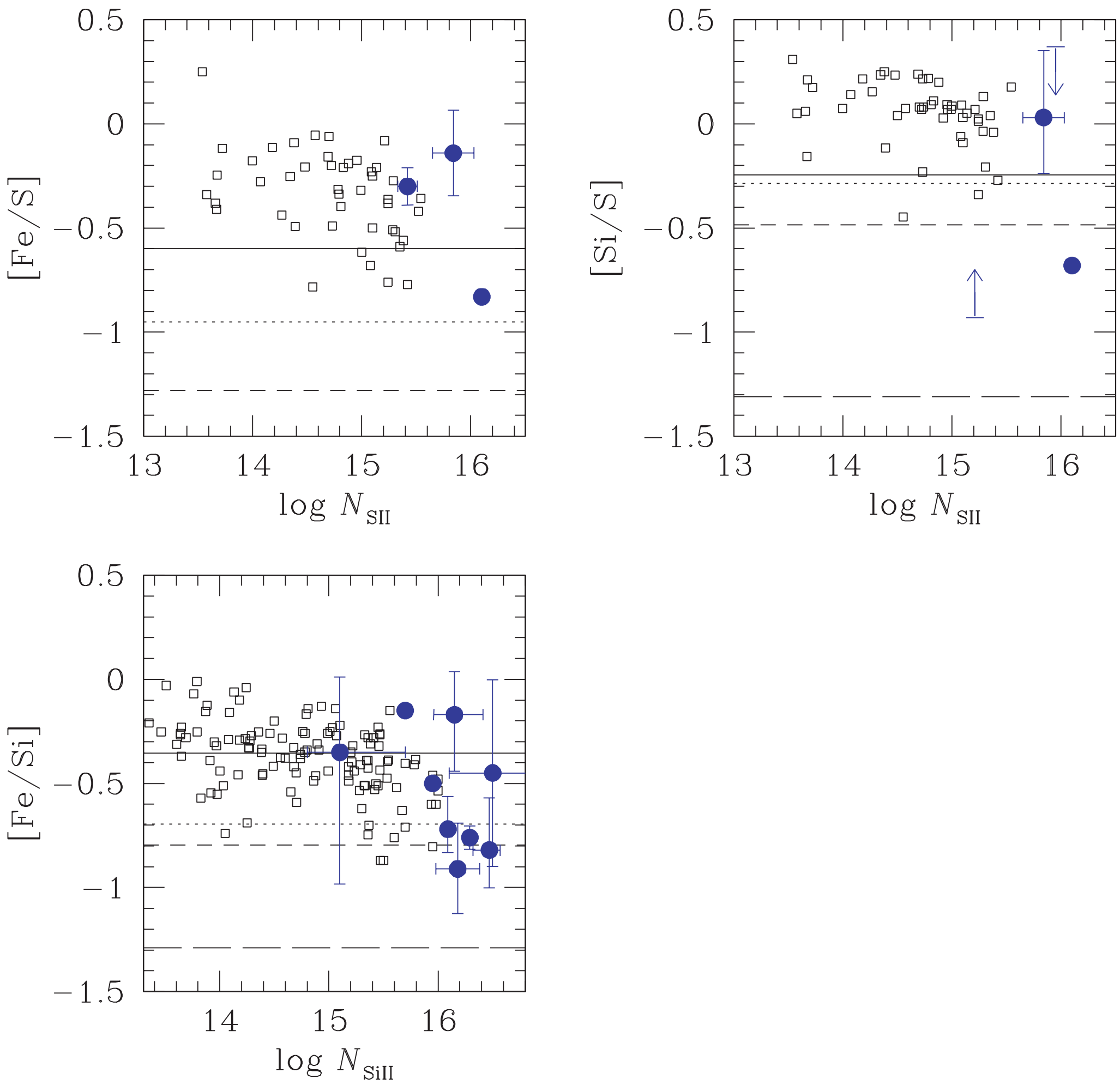

**Figure 6.** Iron-to-sulfur (upper-left panel) and silicon-to-sulfur (right-hand side panel) relative abundances as a function of SII column density, and iron-to-silicon (bottom-left panel) relative abundances as a function of SiII column density. (□) and (●) (and arrows) are QSO-DLAs and GRB-DLAs, respectively. Horizontal lines are mean values derived in the warm halo (solid lines), warm halo + disc (dotted lines), warm disc (short-dashed lines) and cool disc (long-dashed line) ISM of the Milky Way [52].

As of today, unfortunately, it has not been possible to detect in detail the GRB-DLA and the host for the same target. The sets of absorption lines in the UV and emission lines in the optical have little overlapping in wavelength, making this task particularly hard.

Emission line fluxes are widely used to determine the chemical state in the warm ISM ($T \sim 5000$–$10\,000$ K) of local [72, 77] as well as high redshift galaxies [19, 20], [78]–[80]. The most common, and also less expensive method used to derive metallicity is through the $R_{23}$ parameter [79, 81] which is a combination of the [OII], [OIII] and H? line fluxes, and gives the oxygen abundance relative to hydrogen. This parameter can be measured in the optical up to $z \sim 1$. Other methods, such as O3N2 and the electron temperature, can give more reliable results (see [82] and references therein). However, these can only be measured in the optical for relatively nearby and bright objects ($z < 0.5$), whereas most identified GRB hosts are at larger redshift.[7]

**Table 3.**

| GRB | $z$ | Ref | $12 + \log(O/H)$ | $\log Z/Z_{\odot}$ |
|---|---|---|---|---|
| 980425 | 0.0085 | [86] | 8.53 | −0.16 |
| 980703 | 0.966 | [87] | 8.56 | −0.13 |
| 990712 | 0.434 | [88] | 8.47 | −0.22 |
| 991208 | 0.706 | [89] | 9.03 | 0.34 |
| 010921 | 0.451 | [90] | 8.3 | ∼−0.4 |
| 011121 | 0.362 | [91] | 8.85 | 0.16 |
| 020405 | 0.691 | [34] | 8.38 | −0.31 |
| 020903 | 0.251 | [92] | 8.37 | −0.32 |
| 030528 | 0.782 | [93] | 8.44 | −0.25 |
| 031203 | 0.1055 | [84] | 8.49 | −0.20 |
| 051221 | 0.5459 | [94] | 8.78 | −0.09 |

The $R_{23}$ parameter is affected by systematic errors that tend to overestimate metallicity. These can be as large as a factor of 1.5 (0.2 dex), and are more severe for higher metallicities [82, 83]. The metallicity was measured in several GRB hosts [84]–[86]. As several calibrators were used by different authors, we recalculated the metallicity using the same $R_{23}$ calibrator only for consistency, and we added a few hosts for which metallicity was never measured. An LMC dust extinction with $A_V = 1$ and some stellar absorption corrections are assumed. However, results are not strongly affected by a different choice of these two parameters (within limits).

The metallicities of 11 GRB hosts are reported in table 3, covering the redshift range $0.01 < z < 1.0$. The average value is 0.7 times the solar metallicity and the 1σ dispersion relatively high: $< 12 + \log(O/H) > = 8.56 \pm 0.23$,[8] median redshift $z = 0.45$.

This mean metallicity is shown in figure 7 as a function of the Hubble time ($z = 0.45$ corresponds to 8.8 Gyr). We compare it with the GRB-DLA metallicities. One can ask whether the host metallicities are expected to be different than those derived in the GRB-DLAs. Several effects can play a role, like systematic effects in the $R_{23}$ measurement which tend to overestimate metallicity, or α-element enhancement which would give a higher metallicity (measured from oxygen) in GRB hosts. Unfortunately, the metallicity was never derived using both emission as well as absorption lines in the same GRB.

We can compare metallicities with the predictions of an empirical model that gives the metallicity of galaxies, for different galaxy stellar masses, as a function of the Hubble time [19]. This is derived from the mass-metallicity ($M$–$Z$) relation, recently measured for local normal galaxies [77], and its evolution with redshifts [19, 20]. According to this model, the galaxy metallicity evolution is described by:

$$\begin{aligned} 12 + \log(O/H) = &- 7.59 + 2.53 \log M_* - 0.0965 \log^2 M_* + 5.17 \log t_H - 0.394 \log^2 t_H \\ &- 0.403 \log t_H \log M_* \end{aligned} \quad (8)$$

[7] See the GRB Host Studies archive at the URL http://www.pha.jhu.edu/~ savaglio/ghosts.

[8] The oxygen abundance is used as a proxy for metallicity. We assumed the solar oxygen abundance provided by [95], which is $12 + \log(O/H)_{\odot} = 8.69 \pm 0.05$.

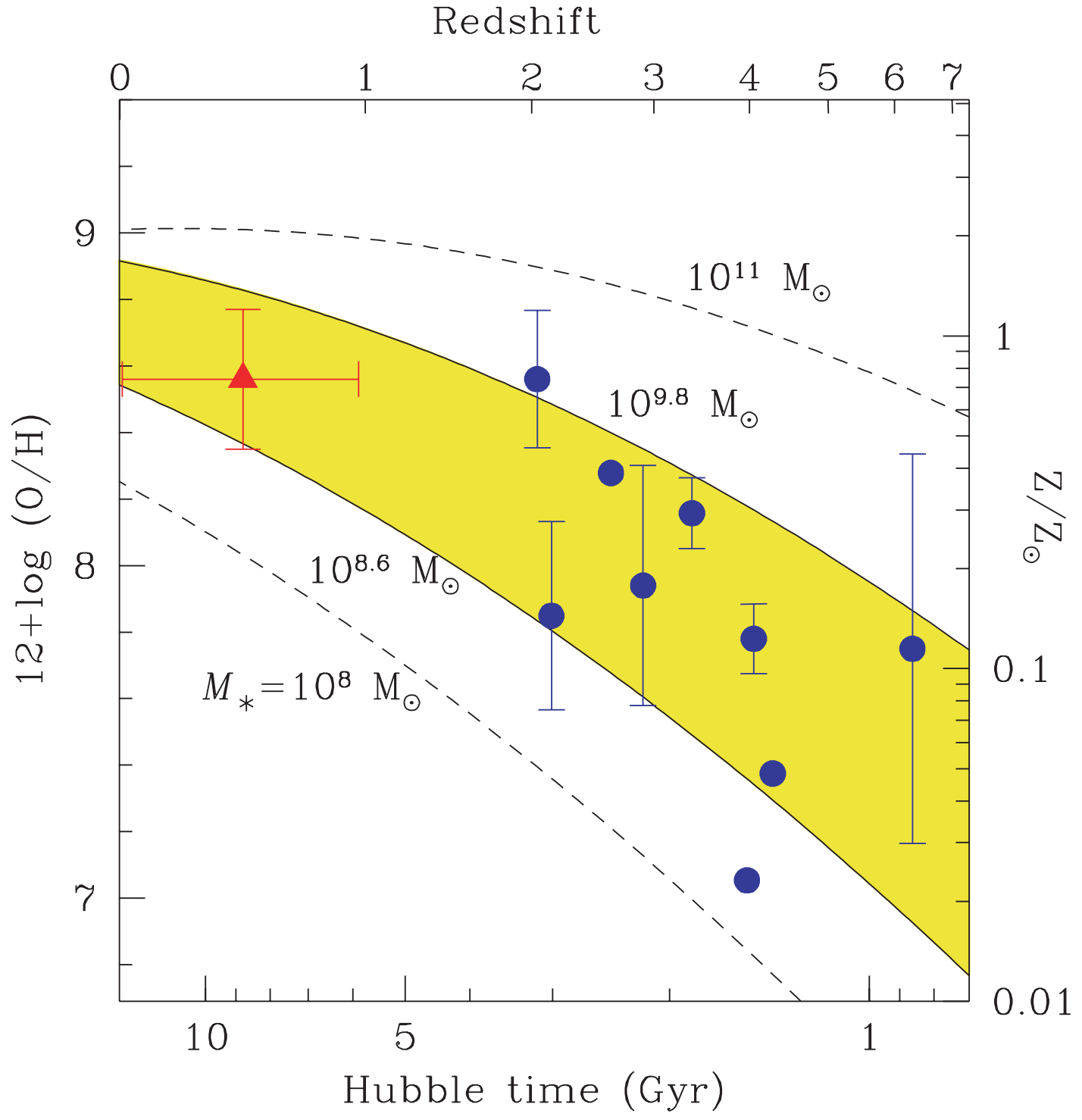

**Figure 7.** Metallicity as a function of Hubble time (lower *x*-axis) or redshift (upper *x*-axis) derived for GRB-DLAs (● —as in figure 5), and the mean value derived for a sample of $z < 1$ GRB hosts (▲, vertical and horizontal bars are the $1\sigma$ dispersion and the redshift interval of the sample, respectively). The curves are predictions from the empirical model of [19], for different total stellar masses. The shaded area indicates the range of stellar masses more favourable for the observed metallicities.

where $t_H$ is the Hubble time in Gyr and $M_*$ is the galaxy stellar mass in solar units. The results of equation (8) are shown in figure 7. Considering the observed metallicities in the GRB hosts and GRB-DLAs, we predict that the stellar mass of the hosts is preferably in the range $M_* = 10^{8.6-9.8} M_\odot$ (shaded area in the figure).

Richer statistics would better support any conclusion. Nonetheless, the predicted low stellar masses are not surprising, because GRB hosts are generally faint [13, 14]. The stellar mass of a sample of 36 GRB hosts in the redshift interval $0.0 < z < 2.6$ was estimated using the observed optical–NIR photometry [21]. The median total stellar mass is $10^{9.1} M_\odot$, with 67% of the masses being in the predicted range of figure 7, i.e. $M_* = 10^{8.6-9.8} M_\odot$.

Our result suggests a consistent picture where the stellar mass of GRB hosts is generally low, and for which metallicities are as expected. Such a picture also indicates that the GRB does not preferentially occur in metal-rich or metal-poor regions, but in regions where metallicity is linked to the stellar mass of the host. GRBs are detected regardless of the host galaxy. Since faint galaxies are the most common in the universe, at low or high redshift, a typical GRB host is more likely a faint galaxy than a bright one. A similar effect is present in galaxies hosting another very energetic event: type Ia supernovae. Type Ia SNe occur in binary systems in which one of the two stars is a white dwarf, with no preference for metallicity. From a sample of more than 100 SNe at $0.2 < z < 0.75$, it was concluded that the SN rate is independent of the host galaxy stellar mass [96]. Yet the typical host stellar mass is low, within the range $M_* = 10^9$–$10^{10.5} M_\odot$ (D Le Borgne, private communication).

A different conclusion for GRB hosts has been recently proposed, according to which there is a relation between the GRB event and the host galaxy: GRB hosts are likely metal-poor and low-mass galaxies because massive metal-poor stars (the GRB progenitors) are rare objects in massive galaxies [73, 74].

## 6. Fine structure

The fine structure of the excited $Si^+$ and $Fe^+$ species was detected in seven GRB afterglows (see table 1 for SiII*). These detections are very useful to study the physical state of the gas [97]. Although the fine structure levels of FeII and SiII seem rather common in GRB afterglow spectra, they were never detected in high-$z$ gas clouds. In QSO-DLAs, only several C fine structure levels were identified [98]. In two recent studies [28, 47] the fine structures of FeII* to FeII****, and SiII* in GRB 050730 and GRB 051111 were discussed in detail. In the latter, additional fine structure transitions of CII and OI are detected [43].

Several scenarios for the physical state of the gas are discussed, but the UV pumping by the GRB is considered the most likely explanation for the observed abundance ratios [28]. Such a picture leads to the conclusion that the gas from which the fine structure originates is at a distance of a few hundreds of pc from the GRB. Due to the fast decline of the GRB emission, a temporal variability of the gas ionization level is expected [99]. A tentative detection of a temporal change in the fine structure line of FeII in the spectrum of GRB 020813 [36, 64] has been recently reported [100].

These gas clouds, though not in the immediate proximity of the GRB, are still part of the GRB environment, in the circumburst medium. This is a small part ($\sim 1/10$) of the total path crossed by the GRB afterglow emission and regularly detected from Earth as a GRB-DLA. From the viewpoint of the heavy element enrichment, these clouds do not strongly affect the total budget of the GRB-DLA metallicity.

At this time, it is premature to add other speculations; further observations will certainly be available in the near future.

## 7. Possible systematic effects

Although GRB science has produced an impressive number of publications, the significance of the results discussed here and in similar papers are still heavily affected by the poor number of statistics. Here, we briefly discuss a few systematic effects that can compromise the general conclusions, in particular on the metallicity derived for GRB-DLAs.

### *7.1. A testbed for the COG: the low-mass high-z starforming galaxy cB58*

Many of the metal column densities discussed in this paper have been estimated using the COG in low–medium resolution spectra. This method has been recently questioned as being inappropriate for measuring column densities [50]. We support the general idea that the COG is a sensitive tool that, if blindly applied, can lead to erroneous results.

This issue is still a matter of controversy, and the best way to clarify it is to use different methods over a sample of high resolution and S/N spectra. The test applied to the spectrum of GRB 051111 suggests that the COG underestimates heavy element column densities [50]. Here, we present another test, performed on the high S/N and moderate resolution (FWHM = 58 km s$^{-1}$) spectrum of the galaxy cB58 at $z = 2.72$, obtained at the Keck II telescope with the spectrograph ESI [51].

Although no GRB was ever observed in this galaxy, it could represent a typical GRB host for several reasons. Its spectrum shows large column densities of HI ($N_{HI} = 10^{20.85}$ cm$^{-2}$) and heavy elements, which translate into a relatively high metallicity of ∼2/5 solar [51], as in GRB-DLAs (figure 7). The fine structure lines of the singly-ionized silicon SiII$^?$ are identified, typical for regions with massive stars [101]. GRB-DLAs, together with cB58, are the only objects outside of the Milky Way where SiII* is detected in absorption (see section 6). As often found for GRB hosts, the star formation rate of cB58 is high, 40–80$M_\odot$ yr$^{-1}$, and its spectrum shows the P Cygni profile, originating in strong winds of massive (> 25 solar masses) and hot stars (Wolf-Rayet stars). A Wolf-Rayet star is considered to be the typical progenitor of long-duration GRBs, as the P Cygni profile was detected in some afterglows and host galaxies [41, 86], [102]–[104]. Given its SFR and age (the starforming episode is younger than 100 Myr), the stellar mass of cB58 is estimated to be relatively low, of the order of $10^{10} M_\odot$ [101], as also seen in GRB hosts [21]. Although not intrinsically bright, cB58 was easily identified as it is gravitationally magnified by an intervening galaxy cluster [105].

The column densities in the ISM of cB58 were carefully derived using the AOD by Pettini *et al* [51]. Here, we follow the COG approach (see also [106]), similarly to what is done for several of the GRB-DLAs reported in table 1. Results derived with COG and AOD are remarkably consistent (figure 8). Although this does not prove that the COG analysis is always correct, it supports the idea that, if properly used, it is reliable. We also note that errors derived from the COG are generally not small (table 1).

The conclusion of Prochaska [50] is that the metallicity in GRB-DLAs is generally underestimated. If this problem affects the measurements of table 2, the difference with QSO-DLAs is even larger than what is suggested by figure 5. Given the estimated stellar masses of the hosts, it is hard to justify very high metallicities unless the starforming region in which a GRB occurs probes higher metallicities than the average host galaxy ambient.

In a recent study [107], column densities of two gas-rich QSO-DLAs obtained from high-resolution (FWHM $\simeq$ 7 km s$^{-1}$) VLT and Keck spectra were compared with the same estimates from medium-resolution (FWHM $\simeq$ 80 km s$^{-1}$) MMT data [108] and results are fully consistent.

### *7.2. Wrong assumption on the dust content*

As discussed in section 3, relative abundances for several GRB-DLAs indicate that the gas is depleted of the most refractory elements. If this is not correct, i.e. if the dust content is negligible, our metallicity measurement, derived using a dust depletion correction, is for many GRB-DLAs incorrect. This would mean that relative abundances are very different than in the Sun's vicinity, which is hard to explain.

The lack of dust depletion would alleviate the problem of the lack of reddening often indicated by the observations of the afterglow spectra [7, 27, 43, 64].

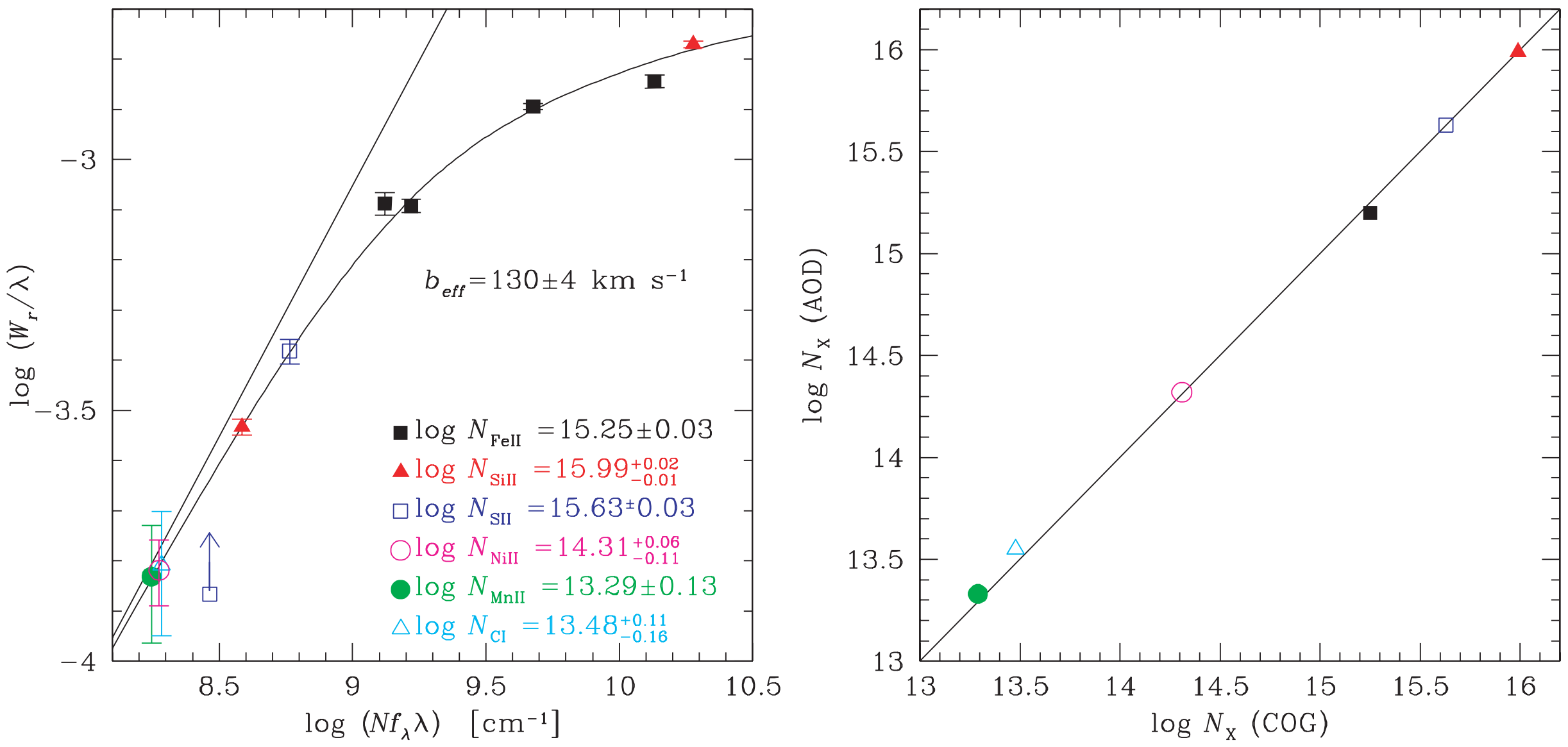

**Figure 8.** Metal column densities derived for the high redshift well-known starforming galaxy cB58 ($z = 2.723$) using standard procedures. On the left-hand side, the COG is used. The effective Doppler parameter is measured from the FeII lines, and is assumed to be the same for the other ions. On the right-hand side, the results are compared with the AOD method used by Pettini *et al* [51]. The typical statistical error is in this case 0.1 dex. Though a GRB was never detected in cB58, it is an archetype GRB host galaxy, characterized by a young stellar population, a high star formation rate and subsolar metallicity. The right-hand side panel shows the consistency of the results derived using the two methods.

On the other hand, even if the dust content is overestimated, and no dust correction is necessary, a difference with the metallicity given by QSO-DLAs is still visible [8, 41]. The general conclusions that GRB-DLAs are extreme objects in the sky, never detected before, and that in general their metallicity is higher than in the typical neutral ISM detected in other high-z galaxies are still valid.

### *7.3. Wrong assumption on the ionization correction*

We neglected the ionization correction in the gas. This assumption is generally justified by the typically high column densities of ions with ionization potential above 13.6 eV, such as FeII, SiII and MgII [24], and by the low abundance of high-ionized species, such as CIV and SiIV [27]. Neutral heavy element species like MgI are often detected in GRB afterglows. However, column densities are generally a small correction with respect to the dominant ions [27, 28].

### *7.4. Possible presence of molecular hydrogen*

The presence of molecular clouds would strongly affect our interpretation of heavy element abundances in GRB-DLAs, if this is a large fraction of the total hydrogen [109]. A few attempts to detect $H_2$ in GRB-DLAs have failed or have given a very low molecular fraction $f_{H_2} = 2N_{H_2}/(2N_{H_2} + N_{HI}) \lesssim 10^{-3.5}$ [7, 8, 41]. Hence, it seems that molecular gas does not affect gas abundances in the GRBs observed so far. It remains to be explained why a large dust content, suggested in section 3, is not producing large molecular clouds. A clumpy distribution of the molecular gas, together with the dust, could escape detection and solve the problem.

Molecular hydrogen in QSO-DLAs is generally very low, in the range $f_{H_2} = 0.02 - 0.001$, or not detected, with $f_{H_2} < 10^{-4}$ [110].

## 8. Discussion

The last decade has seen an exponential increase in the number of scientific results inspired by the GRB phenomenology. Only 1/3 of all papers discussed in this contribution were published before 2003! Many of these are not even directly related to GRB science. Nonetheless, the number of GRBs investigated is still incredibly low, an order of magnitude lower than the related number of papers. This is amazing, considering that bursts occur daily in the sky. Such a large number of publications is a sign of our limited understanding of their nature.

GRBs, being so bright and detectable at the highest redshifts ever explored, can very effectively be used to measure the chemical enrichment in galaxies, from the local universe to when it was only a few hundred Myr old.

It is rather convincing that the cold neutral gas in the host galaxies at $z > 2$ (detected as a GRB-DLA) is characterized by remarkably large column densities of hydrogen and heavy elements. Observations indicate that the metallicity (heavy elements relative to hydrogen) is higher than in the cold ISM of normal high redshift galaxies. The latter is mainly detected using QSOs as background sources. As GRBs are for some time generally brighter than QSOs, it was suggested that GRBs probe the chemical enrichment of the ISM in high-$z$ galaxies in a more unbiased way. We favour the idea that GRBs are the only possible objects that make the detection of very dense starforming regions at high redshifts an affordable task, given the present technology.

If the cold neutral gas metallicities at $z > 2$ are combined with the metallicities derived from the warm gas in the hosts at lower redshift ($z < 1$), then a metallicity redshift evolution is observed.

The measured metallicities are generally lower than those derived from emission lines in starforming galaxies, detected in local to $z \sim 2$ surveys [19, 20, 77]. If one considers the mass–metallicity relation (according to which more massive galaxies are more metal-rich) and its redshift evolution, we expect the stellar mass of the host galaxies to be likely in the range $M_* = 10^{8.6}$–$10^{9.8} M_\odot$, with a mean value similar to the stellar mass of the LMC.

GRB host galaxies have been found to be fainter than typical galaxies detected by ordinary surveys, which is by itself an indication of low stellar masses. The stellar mass range, derived using optical–NIR photometry in $z < 2$ GRB hosts, was found to be similar to that predicted here using the mass–metallicity relation [21, 111]. The low stellar masses are not surprising because low-mass galaxies are the most numerous in the universe at any redshift. GRB hosts are not special, but just normal, faint, generally starforming galaxies, detected at any redshift just because a GRB event has occurred. Such a statement does not support the conclusion according to which GRB hosts tend to be more metal-poor than normal galaxies with similar masses [73, 74].

Low-mass galaxies are an important component from the galaxy formation standpoint, and they can efficiently be studied using GRBs. In the distant universe, similar galaxies have never been spectroscopically identified using conventional astronomy. A rough comparison of GRB host masses with those of normal high-$z$ galaxies indicates that galaxies with $M_* = 10^{9.2} - 10^{10.6} M_\odot$ are about five times more numerous than those with $M_* > 10^{10.6} M_\odot$ [21]. An interesting strategy is to form a complete population of GRB hosts by investigating every single GRB detected by Swift. During the Swift life time, a precise localization of about 600 GRBs is possible, a significant sample for which meaningful studies can be done.

Conclusions are still affected by poor statistics. Although Swift has enormously improved our potentiality, useful spectra are acquired for only a fraction of the afterglows. Recent programs go in the right direction by planning to target all optically-detected GRBs. Nonetheless, some aspects of this research will not improve with more GRB localizations. The metallicity of GRB-DLAs at $z < 1.5$ cannot be measured, as no UV spectrographs are available at this time. The next UV instrument on HST, the Cosmic Origin Explorer, would serve this aim, but it will be online only in a few years from now. Though HST is small compared to ground-based telescopes, the difference in the collecting area is compensated by the higher apparent luminosity of lower redshift GRBs. COS can acquire a high-resolution ($R = 20\,000$) spectrum with $S/N = 10$ at $\lambda = 1800$ Å in 1 h exposure, for a target with AB magnitude of 19.9. This would have been the magnitude of the famous GRB 990123 [112] one day after the trigger if it were at $z = 0.5$. The main limitation with HST is its response time, which can be longer than 2 days.

Another important and not fully understood aspect of GRB observations is the role played by dust. Observed quantities do not match expectations. Our view is strongly biased towards the knowledge of the local, present-time universe. If dust is very abundant along the GRB sight line, it is unclear why many GRBs show very little reddening. A grey dust extinction is possible if the grain size distribution is skewed towards big grains. Another possibility is a strong contamination by clumpy dust [63] which does not considerably change the spectra slope of the background source. One way to investigate this issue is to search for any correlation between the observed spectral slope of the GRB afterglow and the observed metal and dust content, which is particularly interesting if the intrinsic spectral slope is universal.

GRBs offer the opportunity to observe the ISM in $z > 5$ galaxies. The Subaru spectrum of the $z = 6.3$ GRB 050904 shows that GRBs are already metal-enriched even at these early stages of the universe, and this has many important implications on our understanding of the first collapsed objects [73, 113]. The detection of higher-$z$ bursts would require NIR spectroscopy, which is affordable if the GRB is observed sufficiently early. GRB 050904 was observed more than two nights after the trigger, when its flux had declined by several magnitudes. If its spectrum had been observed the night of the trigger, it would have been much brighter than any QSO at $z > 5$, and would have revealed with unprecedented detail the chemical enrichment in the primordial universe.

Bursts like these must be common. Swift has already detected three at $z \gtrsim 5$, out of the 38 GRBs with measured redshift in the first 1.5 years of operation. If Swift will be as successful for the following 4 years, it can detect ten more (see [114] for a more detailed estimate). Future dedicated projects can help to increase this number. One of these is the forthcoming GROND (Gamma-Ray Burst Optical/Near-Infrared Detector), which will provide for the first time simultaneous followups in four optical and three NIR bands, and give in real time a fairly accurate photometric redshift.

The knowledge of the metal enrichment around GRBs is still very open to diverse speculations. Those outlined here are still indicative, yet very exciting. Before the advent of the GRB era, nobody would have ever predicted that nature could have evolved the way GRBs illustrate.

## Acknowledgments

Hsiao-Wen Chen, Mike Fall, Cristiano Guidorzi and Daniele Pierini are acknowledged for inspiring conversations, and Karl Glazebrook and Damien Le Borgne for their important contribution. I am very grateful to Uta Grothkopf for carefully reading the paper, and I also acknowledge kind support from the Physics and Astronomy Department of the Johns Hopkins University, where the initial part of this work was done.